\documentclass{elsart}

\usepackage{graphicx}
\usepackage{dcolumn}
\usepackage{bm}
\usepackage{amssymb}

\newcommand{\bra}[1]{\left(#1\right)}

\begin{document}

\begin{frontmatter}

\title{Transient Dynamics of Sparsely Connected Hopfield Neural Networks with Arbitrary Degree Distributions} \thanks{Physica A 387, 1009(2008)}
\author{Pan Zhang and Yong Chen\corauthref{cor1}}
\corauth[cor1]{Corresponding author. Email address: \tt {ychen@lzu.edu.cn}}
\address{Institute of Theoretical Physics, Lanzhou University, Lanzhou 730000, China}

\date{\today}

\begin{abstract}
Using probabilistic approach, the transient dynamics of sparsely connected Hopfield neural networks is studied for arbitrary degree distributions. A recursive scheme is developed to determine the time evolution of overlap parameters. As illustrative examples, the explicit calculations of dynamics for networks with binomial, power-law, and uniform degree distribution are performed. The results are good agreement with the extensive numerical simulations. It indicates that with the same average degree, there is a gradual improvement of network performance with increasing sharpness of its degree distribution, and the most efficient degree distribution for global storage of patterns is the delta function.
\end{abstract}

\begin{keyword}
neural networks \sep complex networks \sep degree distribution \sep probability theory
\PACS 87.10.+e \sep 89.75.Fb \sep 87.18.Sn \sep 02.50.-r
\end{keyword}

\end{frontmatter}


As a tractable toy model of associative memories and can also be viewed as an extension of the Ising model, Hopfield neural  networks \cite{Hopfield1982} received lots of attention in recent two decades. Equilibrium properties of fully-connected Hopfield neural network have been well studied using spin-glass theory, especially the replica method \cite{Amit85a,Amit85b}. Dynamics is also studied using generating functional method \cite{Coolen2000} and signal-to-noise analysis \cite{Amari88,Okada95,Bolle03} .

Given the huge number of neurons, there is only small number of interconnections in human brain cortex ($\sim 10^{11}$ neurons and $\sim 10^{14}$ synapses). In order to simulate a biological genuine model rather than the fully-connected networks, various random diluted models were studied, including extremely diluted model \cite{DGZ87,PZ90}, finite diluted model \cite{Theumann03,ZC06}, and finite connection model \cite{WC03,CS04}. But neural connectivity is suggested to be far more complex than fully random graph, e.g. the networks of c.elegans and cat's cortical neural were reported to be small-world and scale-free, respectively \cite{nncomplex1,Watts98}. To go one step closer to more biological realistic model, many numerical studies are carried out, focusing on how the topology, the degree distribution, and clustering coefficient of a network topology affect the computational performance of the Hopfield model \cite{ComplexNN,Patrick03,Torres03,Kim04}. With the same average connection, random network was reported to be more efficient for storage and retrieval of patterns than either small-world network or regular network \cite{Patrick03}. Torres \emph{et al.} reported that the capacity of storage is higher for neural network with scale-free topology than for highly random diluted Hopfield networks \cite{Torres03}. However, to our best knowledge, there are no any theoretical results of either dynamics or statics yet.

The goal of this paper is to analytically study the dynamics of Hopfield model for a sparsely connected topology whose degree distribution is not restricted to a specific distribution (e.g. Poisson) but can take arbitrary forms. Another question investigated in this paper is how the degree distribution of connection topology influences the network performance, especially whether there exists an optimal degree distribution given a fixed number of nodes and connections.

Let us consider a system of $N$ spins or neurons, the state of the spins takes $s_i\left(t\right)=\pm 1$ and updates synchronously with the following probability,
\begin{equation} \label{eq-01}
\mathrm{Prob} [s_i \left(t+1\right) | h_i\left(t\right) ] = \frac{e^{\beta s_i\bra{t+1}h_i\bra{t}}}{2 \cosh \bra{\beta h_i\bra{t}}},
\end{equation}
where $\beta$ is the inverse temperature and the local field of neuron $i$ is defined by
\begin{equation}\label{eq-02}
h_i\left(t\right)=\sum_{j=1}^NJ_{ij}s_j\left(t\right).
\end{equation}
We store $q = \alpha N$ random patterns $\xi^ \mu = \left( \xi_1^\mu,\ldots,\xi_N^\mu \right)$ in networks, where $\alpha$ is called the loading ratio. The couplings are given by the Hebb rule,
\begin{equation}\label{eq-03}
J_{ij} = \frac{C_{ij}} {N} \sum _{\mu=1}^q \xi_i^{\mu} \xi_j^\mu,
\end{equation}
where $C_{ij}$ is the adjacency matrix ($C_{ij}=1$ if $j$ is connected to $i$, $C_{ij}=0$ otherwise). In contrast to spin glasses or many other physical systems, the interactions between biological neurons are not symmetric: neuron $i$ may influence neuron $j$ even if neuron $j$ has no influence on neuron $i$. So in our model, $C_{ij}$ and $C_{ji}$ are chosen independently. Degree of spin $i$, $k_i = \sum _{j=1}^N C_{ij}$, denotes the number of spins that are connected to $i$. We consider the case that neurons are sparsely connected, it means that $N\rightarrow \infty$, $k_i\rightarrow \infty$ but $k_i/N\rightarrow 0$. For example, we can take $k_i=O(\ln N)$. And in this paper, the degrees of neurons are set as an arbitrary distribution $p\bra{k_i=k}$.

We use $g\left(\cdot\right)$ to express the transfer function,
\begin{equation}\label{eq-04}
s_i\left( t+1 \right) = g\left( h_i(t) \right).
\end{equation}
Without loss of generality, let us consider the case to retrieve $\xi^1$. We define $m\bra{t}$ as the overlap parameter between network state $s\bra{t}$ and the first pattern $\xi^1$ as
\begin{equation}\label{eq:overlap:def}
    m\bra{t} = \frac{1}{N} \sum_{i=1}^N \xi_i^1 s_i\bra{t}.
\end{equation}
Then the local field at time $t$ can be represented by
\begin{equation}\label{eq:h1}
h_{i} \bra{t} = \frac{1}{N} \sum_{j\neq i}^N C_{ij} \xi_{i}^{1} \xi_j^1 s_j\left(t\right) + \frac{1}{N} \sum_{\mu\neq 1}^q \sum_{j\neq i}^N C_{ij} \xi_i^{\mu} \xi_j^\mu s_j\left(t\right),
\end{equation}
where the first term is the signal from $\xi^1$ and the second one is crosstalk noise from other patterns. Our aim is to determine the form of the local field in the thermodynamic limit $N\rightarrow\infty$. We apply the law of large numbers to the signal term and find that it converges to $\xi_i^1 \frac{k_i}{N} m\bra{t}$ in the thermodynamic limit. To show this point intuitively, we can simply replace the signal term by its average,
\begin{equation}\label{eq:m:converge:1}
    \lim_{N\rightarrow\infty} \frac{1}{N} \sum_{j\neq i} C_{ij} \xi_{i}^{1} \xi_j^1 s_j\left(t\right) {=}\xi_{i}^{1} \left< C_{ij}  \xi_j^1 s_j\left(t\right) \right>.
\end{equation}
This formula is exact in the thermodynamic limit because the whole system is assumed to be self-averaging. Since $C_{ij}$ and $\xi^1$ are independent of each other, we can write the average of product as the product of average,
\begin{equation}\label{eq:m:converge:2}
    \xi_{i}^{1} \left< C_{ij} \xi_j^1 s_j \left(t\right) \right > = \xi_{i}^{1} \left< C_{ij} \right> \left< \xi_j^1 s_j \left(t\right) \right>.
\end{equation}
Using definition of $k_i$ together with Eq.~(\ref{eq:overlap:def}), we have $\left< C_{ij} \right> = k_i/N$ and $m\bra{t} = \left< \xi_j^1 s_j \left(t\right) \right>$. So we have following formula,
\begin{equation}\label{eq:m:converge}
    \lim_{N\rightarrow\infty} \frac{1}{N} \sum_{j\neq i} C_{ij} \xi_{i}^{1} \xi_j^1 s_j\left(t\right) = \xi_i^1 \frac{k_i}{N} m\bra{t}.
\end{equation}

Taking a closer look at the second term in Eq.~(\ref{eq:h1}), if all the terms in the sum (with regard to $\mu$) are independent, we are able to apply the central limit theorem to it. As pointed out in Ref.~\cite{DGZ87}, two conditions are essential for the independence of terms in the sum: first is that almost all feedback loops are eliminated, and the second is that with probability $1$, any two neurons have different clusters of ancestors, i.e. they will remain independent because they receive inputs from two trees which have no neurons in common. In our model, because of the sparsely connected architecture together with the high asymmetry of synaptic connections, two conditions are both satisfied. Thus the second term in Eq.~(\ref{eq:h1}) converges to a zero-mean Gaussian form $\mathcal{N}\bra{0, \frac{\bra{q-1}k_i}{N^2}}$ where $\frac{\bra{q-1}k_i}{N^2}$ is the variance of Gaussian noise. Then the local field of neuron $i$ can be expressed by
\begin{equation}\label{eq:h2}
    h_i\bra{t} = \xi_i^1 \frac{k_i}{N} m\bra{t} + \mathcal{N} \bra{0, \frac{\bra{q-1}k_i}{N^2}}.
\end{equation}
Note that similar treatment of local field can also be found in \cite{Bolle03}.

Then the average state of neuron $i$ was given formally by
\begin{equation}\label{eq:average:state}
    \left< s_i\bra{t+1} \right> = \int dz \bra{2\pi}^{-1/2} \exp \bra{-z^2/2} \left< g\bra{\xi^1 \frac{k_i}{N} m\bra{t} + \frac{\sqrt{\bra{q-1}k_i}}{N} z} \right>_{\xi^1},
\end{equation}
where $\langle \rangle _{\xi^1}$ stands for averaging over distribution of $\xi^1_i$, and $P(\xi) = \left[ \delta(\xi+1) + \delta(\xi - 1) \right ] /2$. When self-averaging is assumed, the average of neuron state in the next time can be obtained by taking average over all $N$ neurons,
\begin{equation}\label{eq:h3}
    \left<s\bra{t+1}\right>=\frac{1}{N} \sum_{i=1}^N \int dz\bra{2\pi}^{-1/2} \exp\bra{-z^2/2} \left<g\bra{\xi^1\frac{k_i}{N} m\bra{t} + \frac{\sqrt{\bra{q-1}k_i}}{N}z} \right>_{\xi^1}.
\end{equation}

Using the concept of degree distribution, we only need to take average over the degree distributions as
\begin{equation}\label{eq:h4}
\left<s\bra{t+1}\right> = \int dk p\bra{k}\int dz\bra{2\pi}^{-1/2} \exp\bra{-z^2/2} \left< g\bra{\xi^1 \frac{k}{N} m\bra{t} + \frac{\sqrt{\bra{q-1}k}}{N}z} \right>_{\xi^1}.
\end{equation}
The overlap parameters are obtained in the similar way,
\begin{equation}\label{eq:m}
    m\bra{t+1} = \int dk p\bra{k}\int dz\bra{2\pi}^{-1/2} \exp\bra{-z^2/2} \left< \xi^1 g\bra{\xi^1\frac{k}{N} m\bra{t} + \frac{\sqrt{\bra{q-1}k}}{N}z} \right>_{\xi^1}.
\end{equation}
When focusing on the most interested case of zero temperature ($\beta\rightarrow\infty$), transfer function $g\bra{\cdot}$ is replaced by $\mathrm{sgn} \bra{\cdot}$. From Eq.~(\ref{eq:m}) one gets
\begin{eqnarray}\label{eq:m:1}
m\bra{t+1} & = \int dk p\bra{k} \frac{1}{2}  \left[ \int_{\frac{k}{N} m\bra{t} + \frac{\sqrt{\bra{q-1}k}}{N} z=0}^{z=+\infty} dz \bra{2\pi}^{-1/2} \exp \bra{-z^2/2} \right. \nonumber\\
& - \int_{z=-\infty}^{\frac{k}{N}m\bra{t}+\frac{\sqrt{\bra{q-1}k}}{N}z=0} dz \bra{2\pi}^{-1/2} \exp \bra{-z^2/2} \nonumber \\
& + \int_{z=-\infty}^{-\frac{k}{N}m\bra{t}+\frac{\sqrt{\bra{q-1}k}}{N}z=0} dz \bra{2\pi}^{-1/2} \exp \bra{-z^2/2} \nonumber\\
&\left. - \int_{-\frac{k}{N}m\bra{t}+\frac{\sqrt{\bra{q-1}k}}{N}z=0}^{z=+\infty} dz \bra{2\pi}^{-1/2} \exp \bra{-z^2/2} \right ]
\end{eqnarray}
Then, the last equation can be further simplified to
\begin{equation}\label{eq:m:0:temperature}
    m\bra{t+1} = \int dk p\bra{k} \mathtt{erf} \bra{\frac{m\bra{t}} {\sqrt{\bra{q-1}/k}}},
\end{equation}
where
\begin{equation}\label{eq:erf}
    \mathtt {erf}\bra{u} = \sqrt{2/\pi} \int_0^u \exp \bra{-x^2/2}dx.
\end{equation}
This finishes the Signal-to-Noise derivation of overlap parameter at zero temperature. As long as the degree distribution of network is determined, using Eq.~(\ref{eq:m}-\ref{eq:erf}), one can calculate temporal evolution of overlap parameters up to an arbitrary time step.

\begin{figure}
\includegraphics{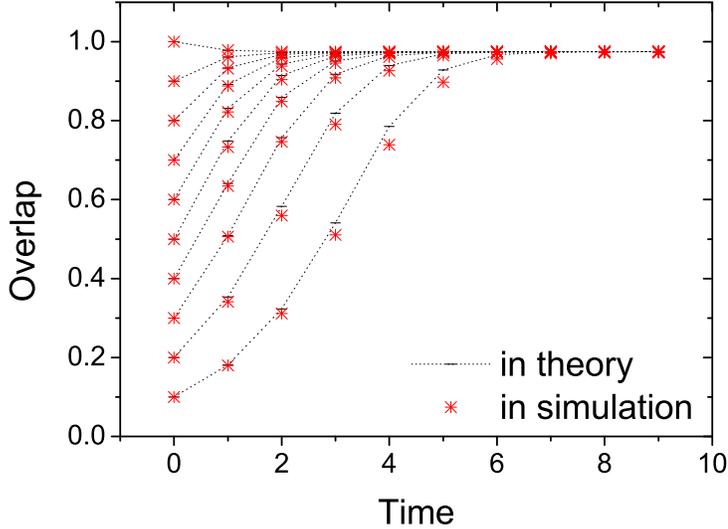}
\caption{Time evolution of overlap parameters for Hopfield network with delta function degree distribution. Initial overlaps range from $1.0$ to $0.1$ (top to bottom). $N$ is $50000$. Each neuron has $100$ degrees and $20$ patterns are stored in networks.} \label{fig:rer}
\end{figure}

Using auxiliary thermal fields $\gamma\bra{t}$ to express the stochastic dynamics \cite{Bolle03}, it is easy to extend the method to arbitrary temperatures by averaging the zero temperature results over the auxiliary fields.
\begin{equation}\label{eq:temperature}
    s\bra{t+1}=g\bra{h\bra{t}+\gamma\bra{t}/\beta},
\end{equation}
and the probability density of $\gamma\bra{t}$ is given by
\begin{equation}\label{eq:noise:dist}
    p\bra{\gamma\bra{t}}=\frac{1}{2} \bra{1-\tanh^2\bra{\gamma\bra{t}}}.
\end{equation}

For illustrative examples, we apply our theory to networks with some specific degree distributions and numerical simulations are performed to verify the theoretical results. In all of our numerical simulations, we set $N = 5 \times 10^4$ and the average degree $\bar{k} = 100$, varying only the arrangement of connections. Each neuron is connected on average to $0.2\%$ of the other neurons compared to $\sim 0.1\%$ in the mouse cortex \cite{BS98}.

The first numerical experiment is the delta function
\begin{equation}
p\bra{k} = \delta \bra{k-\bar{k}},
\end{equation}
which means that every neuron has exactly $\bar{k}$ connections. In practice, the connection topology is generated by randomizing a regular lattice which average degree is $\bar{k}$. Time evolutions of overlap parameters from theory and numerical simulations are plotted in Fig.~\ref{fig:rer}.

\begin{figure}
\includegraphics[width=1\textwidth]{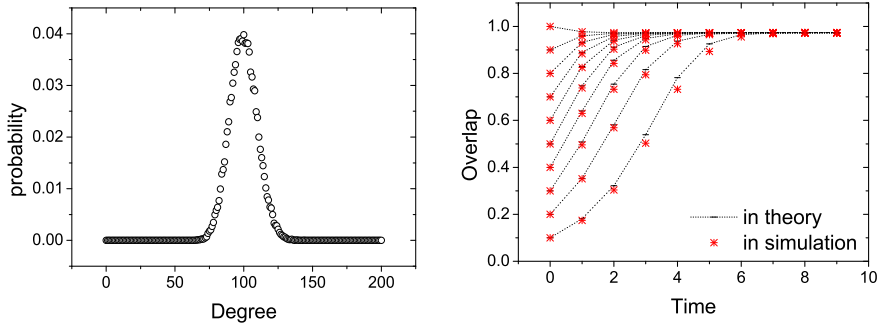}
\caption{Left panel: the normalized binomial degree distribution of networks. Right panel: the temporal evolution of overlap parameters for Hopfield network with degree distribution shown in the left panel (Erd\"{o}s-Renyi random graph). Initial overlaps range from $1.0$ to $0.1$ (top to bottom). $N=50000$, $\bar{k}=100$, and $20$ patterns are stored in networks.} \label{fig:r1}
\end{figure}

The second degree distribution is binomial distribution which comes from a Erd\"{o}s-Renyi random graph \cite{ER} (see the left panel of Fig.~\ref{fig:r1})
\begin{equation}
p\bra{k} = C_N^k \bra{\frac{\bar{k}}{N}}^k \bra{1-\frac{\bar{k}}{N}}^{N-k}.
\end{equation}
The temporal evolution of overlap parameters are presented in the right panel of Fig.~\ref{fig:r1}.

\begin{figure}
\includegraphics[width=1\textwidth]{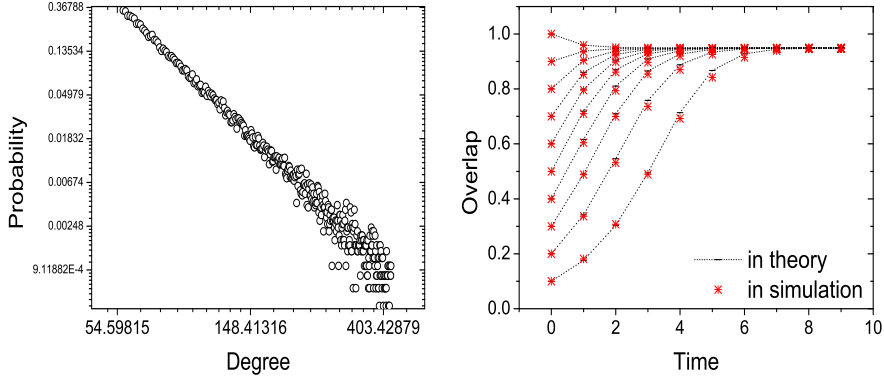}
\caption{Left panel: the normalized power-law degree distribution (log-log scale). Right panel: the time evolution of overlap parameters for Hopfield network with  degree distribution plotted in the left panel. Initial overlaps range from $1.0$ to $0.1$ (top to bottom). $N=50000$, $\bar{k}=100$, and $20$ patterns are stored in networks.}
\label{fig:sfran}
\end{figure}

The third one is power-law distribution (see the left panel of Fig.~\ref{fig:sfran})
\begin{equation}
p\bra{k}=\frac{1}{2}\bar{k}^2k^{-3},
\end{equation}
which is of great importance because it may comes from preferential attachment in the growth process of neurons \cite{Ba99}. The right panel of Fig.~\ref{fig:sfran} shows the temporal evolutions of overlap parameters.

It is obvious that the theoretical results from our scheme are consistent with the simulations for the above degree distributions. Emerging naturally from the above statements, which form of degree distribution is the best one?

\begin{figure}
\includegraphics{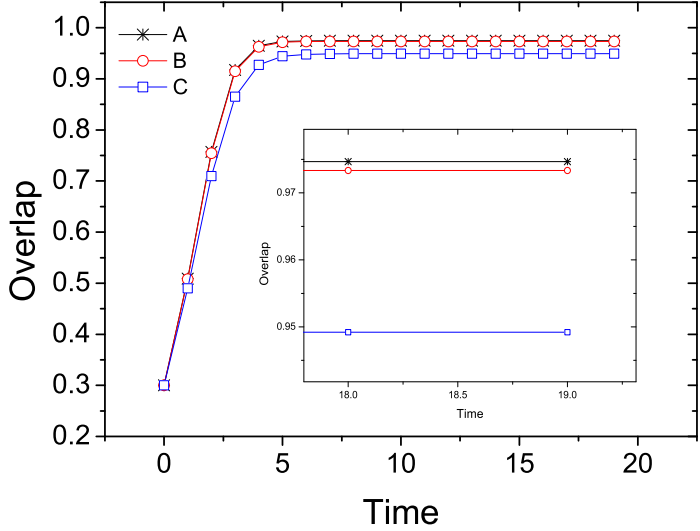}
\caption{Theoretical comparison of time evolutions of overlap parameters in networks with the same average degree but different degree distributions. Degree distribution of $A$ is the delta function, $B$ is binomial, and $C$ is power-law. The inset shows in detail that performance of network with delta function degree distribution is slightly better than that with binomial distribution (Erd\"{o}s-Renyi random graph). $N=50000$, $\bar{k}=100$, and $20$ patterns are stored in networks.} \label{fig:rer-r1-sfran}
\end{figure}

\begin{figure}
\includegraphics[width=1\textwidth]{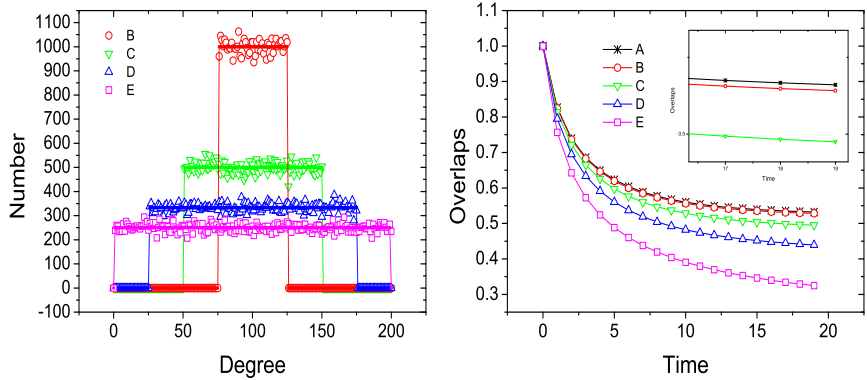}
\caption{Theoretical comparison of time evolutions of overlap parameters for Hopfield networks for uniform degree distributions with different width. Degree distribution of $A$ is delta function. Left panel: the uniform degree distributions with various width, $50$ ($B$), $100$ ($C$), $150$ ($D$), and $200$ ($E$). The inset in right panel shows in detail that the performance of $A$ is slightly better than that of $B$. $N=50000$, $\bar{k}=100$, and $55$ patterns are stored in networks.} \label{fig:uniform}
\end{figure}

To investigate how degree distributions influence the performance of networks, we theoretically compare time evolutions of overlap parameters with delta function ($A$), binomial ($B$), and power-law degree distribution ($C$) in Fig.~\ref{fig:rer-r1-sfran}. The inset shows the details near stationary states. It indicates that network with delta degree distribution performs slightly better than that with binomial distribution. The most rapid degradation in overlap occurs in network with power-law distribution. This behavior can be interpreted as follows. Using Eq.~(\ref{eq:m:0:temperature}), it is easy to find that an individual neuron with fewer degrees suffers more perturbations from crosstalk noise. In the case of power-law degree distribution, degrees are not uniformly distributed in networks and there are too many neurons with small number of connections, which leads to negative performance of the entire networks. However, note that despite the disadvantages of power-law distribution, hubs (subset of networks which has higher degrees) in networks may be useful for partial storage \cite{Patrick03}.

In addition, for verifying the above statements, we construct the special cases that the degree distributions of networks is uniform with various widthes, $50$ ($B$), $100$ ($C$), $150$ ($D$), $200$ ($E$), and the delta function $A$ which width is $0$ (see the left panel of Fig.~\ref{fig:uniform}). The temporal evolutions of overlaps are plotted in the right panel of Fig.~\ref{fig:uniform}, and the inset shows the detailed information near stationary states. It was found that the much more widespread degree distribution tends to induce worse performance of networks.

In summary, the transient dynamics of sparsely connected Hopfield model with arbitrary degree distributions is studied in this paper. It was found that the delta function degree distribution is optimal in terms of network performance, and there is a gradual improvement for network performance with increasing sharpness of its degree distribution. We would like to emphasize that the model investigated in this paper is a simple relaxation to real network topology, by neglecting loops in it. But Ref. \cite{ZC07} suggested that the feedback loops together with their correlations exist in networks and play important role in network dynamics even in the case of sparsely connected systems. It would be interesting to investigate Hopfield model with real complicated topology influenced both by degree distributions and loops (feedbacks and correlations).

\section*{Acknowledgements}
This work was supported by the National Natural Science Foundation of China under Grant No. $10305005$ and by the Special Fund for Doctor Programs at Lanzhou University. One of us (PZ) thanks Dong Liu for useful suggestions.


\begin{thebibliography}{00}

\bibitem{Hopfield1982} Hopfield, J. J. (1982).  Neural Networks and Physical Systems with Emergent Collective Computational Abilities. \textit{Proc. Nat. Acad. Sci. USA}, 79(8), 2554-2558.

\bibitem{Amit85a} Amit, D. J., Gutfreund, H., \& Sompolinsky, H. (1985). Spin-glass models of neural networks. \textit{Phys. Rev. A}, 32(2), 1007-1018.

\bibitem{Amit85b} Amit, D. J., Gutfreund H., \& Sompolinsky, H. (1985). Storing Infinite Numbers of Patterns in a Spin-Glass Model of Neural Networks. \textit{Phys. Rev. Lett.}, 55, 1530-1533.

\bibitem{Coolen2000} Coolen, A. C. C. (2000). Statistical Mechanics of Recurrent Neural Networks II. arXiv:cond-mat/0006011.

\bibitem{Amari88} Amari, S., \& Maginu, K. (1988). Statistical neurodynamics of associative memory. \textit{Neural Networks}, 1, 63-73.

\bibitem{Okada95} Okada, M. (1995). A Hierarchy of Macrodynamical Equations for Associative Memory. \textit{Neural Networks}, 8, 833-838.

\bibitem{Bolle03} Bolle, D., Blanco, J. B., \& Verbeiren T. (2004). The signal-to-noise analysis of the Little-Hopfield model revisited. \textit{J. Phys. A}, 37, 1951-1969.

\bibitem{DGZ87} Derrida, B., Gardner E., \& Zippelius, A. (1987). An exactly solvable asymmetric neural network model. \textit{Europhys. Lett.}, 4, 167-173.

\bibitem{PZ90} Patrick, A. E., \& Zagrebnov, V. A. (1990). Parallel dynamics for an extremely diluted neural network. \textit{J. Phys. A}, 23, L1323-L1337.

\bibitem{Theumann03} Theumann, W. K. (2003). Mean-field dynamics of sequence processing neural networks with finite connectivity. \textit{Physica A}, 328, 1-12.

\bibitem{ZC06} Zhang, P., \& Chen, Y. (2007). Statistical neurodynamics for sequence processing neural networks with finite dilution. \textit{Lect. Note Comput. Sci.}, 4491, 1144-1152.

\bibitem{WC03} Wemmenhove, B., \& Coolen, A. C. C. (2003). Finite connectivity attractor neural networks. \textit{J. Phys. A}, 36, 9617-9633.

\bibitem{CS04} Castillo, I. P., \& Skantzos, N. S. (2004). The Little-Hopfield model on a Random Graph. \textit{J. Phys. A}, 37, 9087-9099.

\bibitem{nncomplex1} Stephan, K.E., Kamper, L., Bozkurt, A., Burns, G.A.P.C., Young, M.P., \& Koter, R. (2001). Advanced database methodology for the collation of connectivity data on the Macaque brain (CoCoMac). \textit{Phil. Trans. R. Soc. Lond. B Biol. Sci.} 356, 1159-1186; Cherniak, C. (1994). Component placement optimization in the brain. \textit{J. Neurosci.} 14, 2418-2427; Scannell, J. W., Burns, G. A. P. C., Hilgetag, C. C., ONeill, M. A., \& Young, M. P. (1999). The Connectional Organization of the Cortico-thalamic System of the Cat. \textit{Cerebral Cortex} 9, 277-299.

\bibitem{Watts98} Watts D. J., \& Strogatz, S. H. (1998). Collective dynamics of 'small-world' networks. \textit{Nature(London)}, 393, 440-442.

\bibitem{ComplexNN} Simard, D., Nadeau L., \& Kr\"{o}gerar, H. (2005). Fastest learning in small world neural networks. \textit{Phys. lett. A} 336(11), 8-15;
Li, C. \& Chen, G. (2003). Stability of a neural network model with small-world connections. \textit{Phys. Rev. E} 68, 052901-4;
Davey, N. \& Adams, R. (2004). High capacity associative memories and connection constraints. \textit{Connection Science} 16(1), 47-65;
Davey N., Christianson, B. \& Adams, R. (2004). High capacity associative memories and small world networks. \textit{Neural Networks Proceedings} 1, 182;
Stauffer, D., Aharony A., da, da Fontoura Costa L., \& Adler, J. (2003). Efficient Hopfield pattern recognition on a scale-free neural network. \textit{Euro. Phys. J. B} 32, 395-399.

\bibitem{Patrick03} McGraw, P. N., \& Menzinger, M. (2003). Topology and computational performance of attractor neural networks. \textit{Phys. Rev. E}, 68, 047102-047105.

\bibitem{Torres03} Torres,  J. J., Munoz, M. A., Marro, J., \& Garrido, P. L. (2003). Influence of topology on the performance of a neural network, arXiv:cond-mat/0310205.

\bibitem{Kim04} Kim, B. J. (2004). Performance of networks of artificial neurons: The role of clustering. \textit{Phys. Rev. E}, 69, 045101-045104.

\bibitem{BS98} Braitenberg, V. \& Sch\"{u}z, A. (1998). \textit{Cortex: Statistics and Geometry of Neruonal Connectivity}. Springer-Verlag, Berlin.

\bibitem{ER} Erd\"{o}s, P., \& Renyi, A. (1959). On random graphs. \textit{Publ. Math. (debrecen)} 6, 290-297.

\bibitem{Ba99} Barab\'{a}si, A. L., \& Albert, R. (1999). Emergence of scaling in random networks, \textit{Science}, 286, 509-512; Barab\'{a}si, A. L., Albert, R., \& Jeong, H. (1999). Mean-field theory for scale-free random networks. \textit{Physica A}, 272, 173-187.

\bibitem{ZC07} Zhang, P., \& Chen Y. (2007). Topology and dynamics of attractor neural networks: the role of loopiness, arxiv:cond-mat/0703405.

\end{thebibliography}
\end{document}